\documentclass{iau} 
\usepackage{graphicx}

\def\ga{\mathrel{\mathchoice {\vcenter{\offinterlineskip\halign{\hfil
$\displaystyle##$\hfil\cr>\cr\sim\cr}}}
{\vcenter{\offinterlineskip\halign{\hfil$\textstyle##$\hfil\cr
>\cr\sim\cr}}}
{\vcenter{\offinterlineskip\halign{\hfil$\scriptstyle##$\hfil\cr
>\cr\sim\cr}}}
{\vcenter{\offinterlineskip\halign{\hfil$\scriptscriptstyle##$\hfil\cr
>\cr\sim\cr}}}}}

\title[IAUS 329.~~Massive stars in clusters] 
{High-mass stars in Milky Way clusters}

\author[Ignacio Negueruela]   
{Ignacio Negueruela$^1$
}

\affiliation{$^1$Departamento de F\'{\i}sica, Ingenier\'{\i}a de Sistemas y
  Teor\'{\i}a de la Se\~{n}al\\ Escuela Polit\'ecnica Superior, Universidad de Alicante\\ Carretera de San Vicente del Raspeig s/n,  E03690, San Vicente del Raspeig, Alicante, Spain\\ email: {\tt ignacio.negueruela@ua.es}}

\pubyear{2017}
\volume{329}  
\jname{The lives and death throes of massive stars}
\editors{J.J. Eldridge, J.C. Bray, L.A.S. McClelland, eds.}
\begin{document}

\maketitle

\begin{abstract}
Young open clusters are our laboratories for studying high-mass star formation and evolution. Unfortunately, the information that they provide is difficult to interpret, and sometimes contradictory. In this contribution, I present a few examples of the uncertainties that we face when confronting observations with theoretical models and our own assumptions.
\end{abstract}

\firstsection 
\section{Introduction}

Young open clusters are the natural laboratories where we can study high-mass star formation and evolution. This is because (most) high-mass stars are born in young open clusters, which provide us with astrophysical context for these objects. Until now, membership in an open cluster has been the only reliable way to estimate the distance to (almost all) high-mass stars. This situation is going to change dramatically in a few months with the release of accurate parallaxes by \textit{Gaia}, but even then open clusters will give us other important parameters, such as age and interstellar extinction, that cannot be easily derived from an isolated star. These proceedings contain many examples of how young open clusters can be used to learn important facts about high-mass star formation and evolution. In this contribution, I will provide a few more examples, with special emphasis on the problems that we face to make effective use of these natural laboratories.

\section{Constraints on high-mass star formation}

By now, there is a general consensus on the idea that the vast majority of high-mass stars are formed in open clusters (\cite{zy07} for a review). There is still, however, a heated debate about an apparently very similar proposition: \textit{all high-mass stars form in clusters}. Going from "\textit{almost all}" to "\textit{all}" is not just a subtlety. Simply because they are very rare, high-mass stars should preferentially be found in the most massive clusters (\cite{elmegreen1983}). However, if the observed initial mass function (IMF) is due to purely stochastical sampling of an underlying distribution, we should ocassionally see a high-mass star form in a low-mass environment (\cite{masch08}). Contrarily, if the processes that form stars have a way of knowing about their environment, we would expect to see tight correlations between the mass of a cloud and the mass of the stars it forms (e.g.\ \cite{weidner2013}). Observations of clusters show that, in general, the mass of the most massive cluster member is a function of the cluster mass (\cite{larson1982,weidner2010}), but the interpretation of observations is compounded by many complicating issues, such as the effects of binarity (\cite{eldridge12}) or even the very definition of open cluster and IMF (\cite{cervi13}). So far, the strongest observational constraint in favour of purely statistical sampling is the detection by \cite[Oey \etal\ (2013)]{oey2013} of a sample of 14 OB stars in the Small Magellanic Cloud (SMC) that meet strong criteria for having formed under extremely sparse star-forming conditions.

This issue is intimately linked to the ongoing discussion about the mechanisms that lead to the formation of very massive stars (very massive star is used here in the general sense of a star significantly more massive than the most numerous high-mass stars, with $M_{*}\sim10$\,--\,$15\:$M$_{\odot}$). At present, there are two main competing theories about how such stars may form: (a) monolithic core accretion, in most respects a scaled-up version of classical low-mass formation theories, where very high opacities allow infalling material to overcome radiation pressure (e.g. \cite{ys02,krumholz2009}) and (b) competitive accretion, where massive stars are formed in cluster cores, benefiting from the gravitational potential of the whole cluster to accrete more material (e.g. \cite{bonbate06,smith2009}). Observations do not provide a definite answer.  On one side, there is (quite) direct evidence of disks around some moderately massive stars (up to $M_{*}\sim25$\,--\,30$\:$M$_{\odot}$; see \cite{beltran16} for a review), supporting the idea of monolithic collapse. On the other hand, massive stars in young clusters seem to lie in the regions of highest stellar density (\cite{rivilla2014}), a key prediction of the competitive accretion scenario that seems at odds with monolithic collapse.

Looking for a different approach, we can explore the environments where we find stars with the earliest spectral types (O2\,--\,O3) in the Milky Way. These objects are not the most massive stars, which generally present WNh spectral types (e.g.\ \cite{schnurr08}), but can safely be assumed to have masses $\ga40\:$M$_{\odot}$ (\cite{massey05}). If we take all the stars with these spectral types in the GOSC v3 (\cite{jesus2013}), we find 20 objects. Of these, 14 lie inside or close to some of the most massive clusters in the Milky Way, but four are in smaller clusters ($M_{\textrm{cl}}\sim10^{3}\:$M$_{\odot}$) and two appear in relative isolation within active star formation regions (\cite{marco17}). Taken at face value, this distribution seems to support the stochastic scenario, but we must exercise caution when interpreting such data.

\begin{figure}[ht!]
\begin{center}
 \includegraphics[width=12cm]{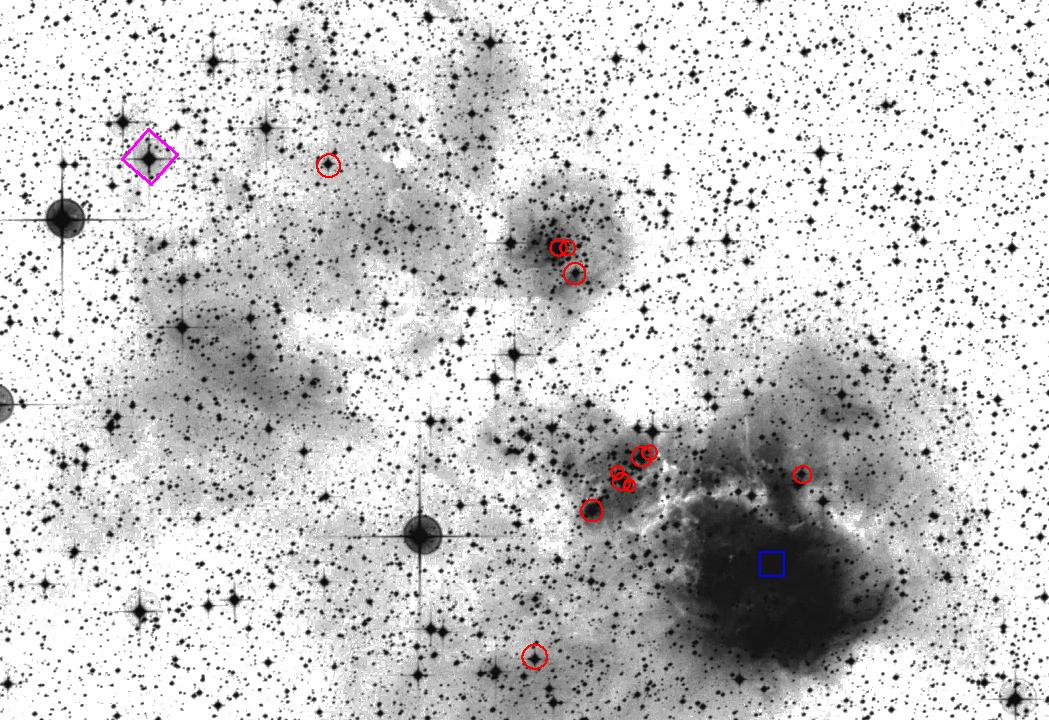} 
 \caption{The position of HD~64\,568 (marked by the purple diamond) relative to the star-forming complex NGC~2467 is shown on this DSS2 blue plate. The position of HD~64\,315 is shown by a blue square. Other OB stars in the area are shown as red circles. The small group to the North is Haffner 18, while the larger concentration towards the centre is Haffner 19.}
   \label{n2467}
\end{center}
\end{figure}

For example, if we take a closer look to one of the two isolated objects, the O3\,V((f*))z star HD 64\,568, in the star-forming complex NGC~2467, we can see that there are almost no nearby OB stars (Fig.~\ref{n2467}). However, it lies about $15^{\prime}$ away from the open cluster Haffner~19, which contains at least one mid-O star. There is some indirect evidence suggesting that it may be a runaway coming from the vicinity of this cluster. Interestingly, to the other side of Haffner~19, the apparently isolated mid-O star HD~64\,315 illuminates the bright H$\alpha$ region Sh2-311. This star was recently classified as O5.5\,Vz+O7\,V (\cite{sota14}), but interferometric observations show that it can be resolved into at least two visual components (\cite{toko10,aldo15}). Analysis of high-resolution spectra shows that the ``star'' includes at least two SB2 systems with a minimum of four O-type stars (\cite{lorenzo17}). Such a complex system cannot have been ejected from a nearby cluster, and must therefore have formed in relative isolation. On the other hand, what was originally thought to be a single star appears now as a stellar aggregate that may well hide a population of lower-mass stars. This hidden multiplicity beautifully illustrates the difficulty in defining isolation or measuring the mass of a small cluster.

\section{Evolution after the main sequence}

Young open clusters provide most of the information that we have about high-mass star evolution. Unfortunately, this information is difficult to interpret, because most young open clusters have quite small populations of (evolved) high-mass stars. As an example, take NGC~457, a typical cluster in the Perseus Arm, with a distance estimate of 2.5~kpc and more than sixty B-type members identified from photometry (\cite{fitz93}, and references therein). This cluster contains three supergiants: the B5\,Ib star HD~7\,902, the F0\,Ia MK standard $\phi$~Cas, and the M2\,Ib red supergiant (RSG) HD~236\,687. With such a complement of evolved stars, it should be an ideal laboratory to constrain the evolution of moderately massive stars. More than $4^{\circ}$ South of the Galactic Plane, and separated from all nearby OB associations, its membership should be essentially free of confusion problems. Abundance determinations for some of its members imply solar-like values for $\alpha$ elements (\cite{dufton94}), allowing direct comparison to theoretical tracks.

\begin{figure}[ht!]
\begin{center}
 \includegraphics[width=8cm]{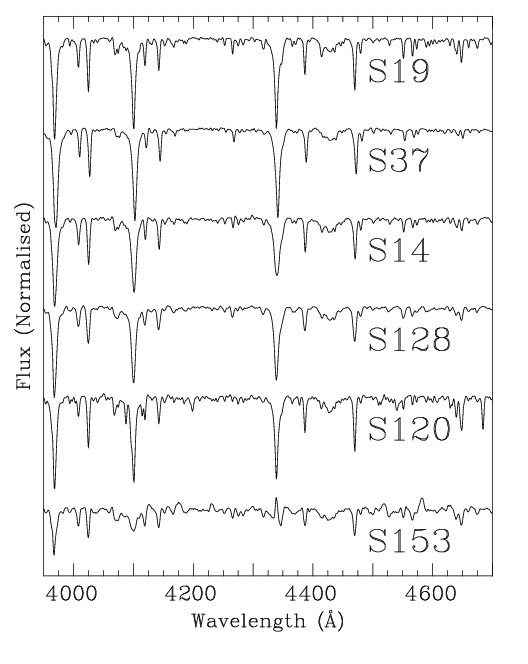} 
 \caption{Classification spectra of six bright blue stars in the open cluster NGC~457. S19, S128 and S37 (and also likely the extreme Be star S153) are all giants between B1.5 and B2. S14 is a very slightly less luminous Be star, while S120 (at O9.7\,V) is a blue straggler. }
   \label{class}
\end{center}
\end{figure}

But gleaning information is not always straightforward. Figure~\ref{class} shows the classification spectra of six of the brightest non-supergiant blue stars in the area of NGC~457. Of them, S153 is an extreme Be star and thus difficult to classify or analyse. Star 19, formally B1.5\,III, looks more luminous than the others, but this is mostly a visual effect due to its very low projected rotational velocity (\cite{huang06}, who also find that S19 and S37 are likely binaries). Star 128 is about as luminous, though a fast rotator. Star 14 is slightly less luminous, and a Be star, while S37 is somewhat later, around B2\,III (and an SB2 binary). The spectral types of these four latter stars are consistent, and their presence at the top of the sequence suggests an age perhaps slightly younger than the $\sim21$~Ma proposed by \cite[Huang \etal\ (2006)]{huang06}. In a Geneva isochrone for $\sim20$~Ma with moderate initial $v_{\textrm{rot}}$ at solar metallicity (\cite{georgy13}), the objects at the top of the sequence have $M_{*}\sim 11\:$M$_{\odot}$. For an age of 16~Ma, we find masses $\sim 13\:$M$_{\odot}$, bracketting the expectations for the spectral type. But S120 is clearly earlier than any other star in the cluster. At O9.7\,V, it should be a blue straggler (as \cite{huang06} measure a radial velocity similar to those of other cluster members). Moving to the supergiants, \cite[Levesque \etal\ (2005)]{levesque05} derive parameters for HD~236\,687 from a global model fit to the spectrum, finding that it is compatible with a $\sim12\:$M$_{\odot}$ track. For HD~7\,902, \cite[McErlean \etal\ (1999)]{mcerlean99}, using tailored model fits, derive parameters compatible with a $\sim 18\:$M$_{\odot}$ star. Finally, the parameters of $\phi$~Cas are much more uncertain. Combining model fits with observational calibrations, \cite[Kovtyukh \etal\ (2012)]{kovtyukh12} derive parameters that would put it well above the $20\:$M$_{\odot}$ track, but \cite[Luck (2014)]{luck14} finds very different values of $\log\, g$ (and, thence, luminosity) when using different atmosphere models. 

Can we use all this apparently conflicting information to constrain stellar evolution? Is there any useful way to mix the results of different groups? All three supergiants have been analysed using different techniques and sets of models. Are the results on the same theoretical scale? Do all groups use the same assumptions? \cite[McErlean \etal\ (1999)]{mcerlean99} use a distance modulus to NGC~457 0.2~mag longer than used by \cite[Levesque \etal\ 2005]{levesque05}. This may not seem like much, and is certainly within the uncertainty of most distance determination methods, but it represents an increase of 10\% in the distance.  Although it cannot explain the higher mass implied for HD~7\,902, it brings it closer to the value derived from the cluster giants.
The position of HD~7\,902 could also perhaps be explained by a very high initial rotational velocity, as stars that rotate very fast become brighter and evolve more slowly (\cite{georgy13}). Another possibility (that could also explain S120) is that it gained mass in a binary while it was still on the main sequence and then followed a standard evolution (\cite{langer}). The properties of $\phi$ Cas could require a more complex explanation. If its mass is $>13\:$M$_{\odot}$, it should not be in a Cepheid loop. Perhaps it is a post-RSG object. There are many possible post-MS tracks, and we only have three stars clearly detached from the main sequence that may have followed very different paths. However, not all hope is lost. There is still a lot of reliable information in the cluster. We observed 15 stars close to the top of the main sequence in NGC~475 (including the six in Fig.~\ref{class}) for the programme described in \cite[Marco \etal\ (2007)]{marco07}. Of these, 12 have spectral types between B1.5 and B2.5, with luminosity classes between V and III, well matched to their relative brightness. The brightest main-sequence stars have spectral type B2\,V providing a good definition of the turn-off, which, through the use of calibrations based on eclipsing binaries, we can place around $9\:$M$_{\odot}$.

\begin{figure}[ht!]
\begin{center}
 \includegraphics[width=12cm]{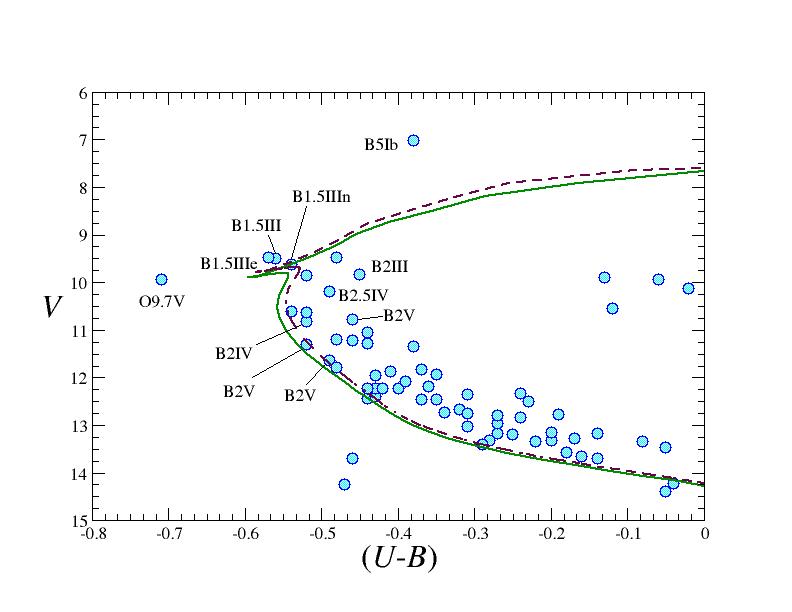} 
 \caption{Photoelectric photometry of blue stars in NGC~457, with our spectral types superimposed. The continuous line is the $16\:$Ma isochrone for no initial rotation, while the dashed line is the $20\:$Ma isochrone for $\Omega$/$\Omega_{\textrm{crit}}=0.3$ from \cite[Georgy \etal\ (2013)]{georgy13}.}
   \label{cmd}
\end{center}
\end{figure}

We can easily check this conclusion. In Fig.~\ref{cmd}, we plot the existing photoelectric photometry for stars in the field of NGC~457, from \cite[Pesch (1959)]{pesch59} and \cite[Hoag \etal\ (1961)]{hoag61}. We can verify the validity of the spectral types, as the photometric turn-off coincides with the transition between class V and higher luminosities. We can then plot the corresponding isochrones from \cite[Georgy \etal\ (2013)]{georgy13}, reddened with the values from \cite[Fitzsimmons (1993)]{fitz93}. We use the 16~Ma isochrone without rotation and the 20~Ma one with $\Omega$/$\Omega_{\textrm{crit}}=0.3$ (they are almost identical). The fit is not perfect, but it is surprisingly good (we just guessed the age based on the spectral types of the brightest stars). A result is much stronger when it is based on many stars hinting at approximately the same conclusion. We can see how the blue straggler lies on the continuation of the main sequence, well to the left of the turn-off. We can also see that the analysis by \cite[McErlean \etal\ (1999)]{mcerlean99} is not in error. The B5\,Ib supergiant is much more luminous than the corresponding isochrone.

To fight low-number statistics, we can resort to utilising information from many different clusters, but then the difficulties increase. The metallicity of young Galactic clusters cannot simply be assumed to be solar. There is a gradient with galactocentric distance and strong local variability (see \cite{negueruela15} and references therein). Moreover, results obtained by different groups can have large systematic effects, affecting many parameters. Membership criteria may vary. Age scales change as new stellar models are developed and successive generations of isochrones are released. For these reasons, the more massive a cluster is (and thus the more high-mass stars it contains) the more useful it is as a laboratory. To continue with our example, the Perseus arm contains a few young open clusters more massive (and hence containing more evolved stars) than NGC~457, namely, NGC~7419, NGC~663, NGC~869, and NGC~884. Their combined analysis shows that blue stragglers happen frequently, but are not very numerous. It also shows that HD~7\,902 is not an exception; blue supergiants (BSGs) almost always fall on tracks rather more massive than the main sequence turnoff (see also \cite{marco07}). These clusters, however, offer very little information about what should be a very strong constraint on evolutionary timescales, the blue to red supergiant ratio (\cite{eggenberger02}), as it varies wildly among them.

\section{Massive clusters}

Clusters with large populations of high-mass stars become much more reliable laboratories. The most extreme example is the starburst cluster Westerlund~1. It contains many dozen supergiants, covering all spectral types at very high luminosities (\cite{clark05}). The sequence of BSGs is well populated from O9, with examples of almost all the B subtypes (\cite{negueruela10}). Under these conditions, the much higher number of O9\,--\,B2 supergiants can be taken as a very strong suggestion of a much faster evolution after $T_{\textrm{eff}}\approx18\,000$~K is reached. Because the population is so large, we find a high number of spectroscopic binaries, including a few eclipsing binaries, which allow a definite measurement of the masses of these stars, which lie in the $\sim35$\,--\,$40\:$M$_{\odot}$ range (\cite{koumpia12}). Moreover, again because of high statistics, we may find binary systems with similar masses, but different orbital parameters. Comparison of their properties lends weight to the concept of diverse evolutionary paths for stars in close binaries (\cite{sana12}). Indeed, in Westerlund~1, we can see examples of many binary systems that represent different stages along these paths (\cite{clark14b})
 
Unfortunately, young massive clusters (YMCs) are rare, and this means that we generally find them at large distances and behind heavy extinction. If we follow \cite[Hanson \etal\ (2015)]{hanson15} in defining a YMC as a cluster with $M_{\textrm{cl}}\ga10^{4}\:$M$_{\odot}$, we know less than twenty YMCs in the Milky Way. Moreover, some of the most massive clusters in the Galaxy, such as NGC~3063 or the Arches cluster, are too young to display any effects of stellar evolution. Given their scarcity, obtaining parameters for YMCs with high accuracy seems paramount, but it is not an easy task. For example, Westerlund~1 is behind more than 10~mag of extinction, and in spite of its astrophysical interest and the effort dedicated over the past ten years, its distance is still uncertain by almost a factor of two.

As an illustration of this difficulty, let us take VdBH~222, one of the less reddened YMCs with $E(B-V)\approx2.5$, despite its location towards the Galactic Centre ($\ell=349^{\circ}$). We observed a large population of red and yellow supergiants, from whose spectra it was easy to measure the radial velocity of the cluster. With a very distinctive value of $v_{\textrm{{\small LSR}}}=-99\pm4\:$km\,s$^{-1}$, the Galactic rotation curve allows two possible distances for VdBH~222, one at $d\sim6$~kpc, and the other at $d\sim11$~kpc (\cite{marco14}). Though such distance difference appears really huge, it only amounts to 1.3~mag in distance modulus. With the high value of $E(B-V)$, a small difference in the value of $R_{V}$ can account for most of this gap. In addition, the shape of isochrones does not change significantly over the age range between $\sim10$ and $\sim20\:$Ma. Only the gap in brightness between the red and blue supergiants can help determining the cluster age, but this also depends on the extinction law. We found that it was possible to fit all photometric data with a $12\:$Ma cluster at $\sim10.5$~kpc or a $18\:$Ma cluster at $\sim6$~kpc (in fact, the photometry allowed many other possibilities between these two, but the radial velocity offers further constraints). The data do not provide any direct way to choose between the two options. However, given our knowledge of Galactic structure, the second option looked rather unlikely, because it places the cluster at less than 3~kpc from the Galactic Centre, in a region supposed to be free of star formation (\cite{marco14}).

However, additional data showed this to be the correct solution. Spectral monitoring of the cool supergiants revealed one of them to be a massive Cepheid. The determination of a 24.3~d period indicated that this is a $\sim10\:$M$_{\odot}$ star (again, the chances of finding such an unusual star increase with cluster population), and so the cluster is definitely on the near side of the Galactic Centre (\cite{clark15}). This has been confirmed by classification spectra of the brightest blue members, which shows all of them to be early-B (bright) giants, and not supergiants as would be required by a higher distance (and hence lower age). Interestingly, this means that this cluster rich in RSGs hosts no BSGs, in analogy to the Perseus arm cluster NGC~7419 (\cite{marco13}). Based on data that we are currently analysing on two other clusters rich in RSGs, this scarcity or even absence of BSGs seems to be a common theme.

Regrettably, obtaining a good characterisation of other clusters rich in RSGs is even more difficult than finding a good solution for VdBH~222, because they are hidden by higher extinction (see a list of these clusters with their parameters in \cite{negueruela16}). In the extreme case of RSGC1, with $A_{K}\approx 2.5$~mag (\cite{davies08}), we are limited to work in the infrared -- even $J$-band spectroscopy is hard to get! But even for the much less obscured Stephenson~2 (\cite{davies07}), we find that spectroscopy in the optical is not feasible, while photometry in the $U$ and $B$ bands is beyond reach. Under this conditions, the reddening law cannot be determined with accuracy, resulting in large uncertainties in the distance to the cluster (and hence all other parameters). Fortunately, there are other tools, and we are likely to see in the near future accurate distances to both Westerlund~1 and Stephenson~2 (and perhaps other clusters) based on trigonometric parallaxes of the SiO masers detected in some of their RSGs (\cite{verheyen12}).  This will allow full characterisation of their properties and a much better use as laboratories of stellar evolution (see \cite{negueruela16} for a discussion of these potentialities).

A great example of this kind of work can be found in the contribution by Beasor \& Davies to these proceedings, where they use the large collection of RSGs in the YMC NGC~2100 (in the LMC) to present strong evidence for an evolutionary sequence among the RSGs: mass-loss rates and luminosity show a clear correlation, as has also been found in an analysis of the LMC field population by \cite[Dorda \etal\ (2016)]{dorda16}. A similar behaviour is readily seen among the red RSG population of Stepheson~2 (\cite{clark14a}). Interestingly, in both clusters the range in brightness of the supergiant population (above 2~mag) is much wider than allowed by any single theoretical isochrone. Why is this? Why do RSGs and BSGs tend to avoid each other in YMCs? Why are BSGs almost always more luminous than the isochrone of their parent cluster? All these questions are presented by the simple observation of young open clusters. Their continued study and accurate characterisation in the \textit{Gaia} era will certainly provide new constraints and new questions.

\acknowledgements
I would like to thank all my collaborators in cluster work, especially Amparo Marco and J. Simon Clark. Hearty thanks to Dr.\ Berto Castro for calculating stellar parameters. Research partially supported by MinECO/FEDER under grant AYA2015-68012-C2-2-P.

 

\end{document}